\documentclass[twocolumn]{aastex62}
\usepackage{graphics,epsf}
\usepackage{amsmath}                
\usepackage{amsfonts}               
\usepackage{amssymb}                
\usepackage{epsfig}                 
\usepackage{graphicx}
\usepackage{float}
\usepackage{color}
\usepackage[para,online,flushleft]{threeparttable}

\newcommand{\erg}{{~\rm erg}}

\newcommand{\days}{{~\rm days}}

\begin{document}

\title{Modelling light curves of bipolar core collapse supernovae from the equatorial plane} 

\author[0000-0003-0375-8987]{Noam Soker}
\affiliation{Department of Physics, Technion, Haifa, 3200003, Israel; noa1kaplan@campus.technion.ac.il; soker@physics.technion.ac.il}
\affiliation{Guangdong Technion Israel Institute of Technology, Shantou 515069, Guangdong Province, China}
 
\author{Noa Kaplan}
\affiliation{Department of Physics, Technion, Haifa, 3200003, Israel; noa1kaplan@campus.technion.ac.il; soker@physics.technion.ac.il}

\begin{abstract}
We use the two-components bipolar toy model of core collapse supernova (CCSN) ejecta to fit the rapid decline from maximum luminosity in the light curve of the type IIb CCSN SN~2018gk (ASASSN-18am). In this toy model we use a template light curve from a different CCSN that is similar to SN~2018gk, but that has no rapid drop in its light curve. The bipolar morphology that we model with a polar ejecta and an equatorial ejecta increases the maximum luminosity and causes a steeper decline for an equatorial observer, relative to a similar spherical explosion. The total energy and mass of our toy model for SN~2018gk are $E_{\rm SN} = 5 \times 10^{51} \erg$ and $M_{\rm SN}= 2.7 M_\odot$. This explosion energy is more than what a neutrino-driven explosion mechanism can supply, implying that jets exploded SN~2018gk. 
These energetic jets likely shaped the ejecta to a bipolar morphology, as our toy model requires. 
We crudely estimate that $f \approx 2-5 \%$ of all CCSNe show this behavior, most being hydrogen deficient (stripped-envelope) CCSNe, as we observe them from the equatorial plane. We estimate the overall fraction of CCSNe that have a pronounced bipolar morphology to be  $f_{\rm bip} \approx 5-15 \%$ of all CCSNe. 
\end{abstract}


\section{Introduction}
\label{sec:intro}
    
The structures of some remnants of core collapse supernovae (CCSNe) and observations indicating polarisation in a number of CCSNe (e.g., \citealt{Wangetal2001, Maundetal2007, Milisavljevic2013, Gonzalezetal2014, Marguttietal2014, Inserraetal2016, Mauerhanetal2017, GrichenerSoker2017, Bearetal2017, Garciaetal2017,  LopezFesen2018})
show the non-spherical explosion nature of at least these CCSNe. 
Some observations suggest the action of jets in at least a fraction of CCSNe, in particular in those remnants that have morphological features similar to those of some planetary nebulae that are shaped by jets (e.g., \citealt{BearSoker2018, Bearetal2017, Akashietal2018}). 
An example is the presence of two opposite protrusions from the main body of a CCSN remnants (`Ears') that resemble some Ears in planetary nebulae where jets are active (e.g., \citealt{Bearetal2017}). Jets with an energy of only $\approx 1 - 10 \%$ of CCSN ejecta energy can account for the presence of Ears \citep{GrichenerSoker2017}. 
To these we can add the recent claim by \cite{Boseetal2019} for a bipolar $^{56}$Ni morphology in the Type II-P CCSN SN~2016X (ASASSN-16at). 
The point to take is that jets can shape the ejecta of some CCSNe to a bipolar structure (e.g., \citealt{Orlandoetal2016, BearSoker2018}).

There are models where only if the pre-collapse core has fast rotation the newly born neutron star (NS) or black hole launch jets (e.g., \citealt{Khokhlovetal1999, Aloyetal2000, Hoflich2001, Burrows2007, Nagakuraetal2011, TakiwakiKotake2011, Lazzati2012, Maedaetal2012, Bromberg_jet, Mostaetal2014, LopezCamaraetal2014, BrombergTchekhovskoy2016, LopezCamaraetal2016, Nishimuraetal2017, Fengetal2018, Gilkis2018}). The demand for a rapidly rotating pre-collapse core makes these mechanisms for jet launching rare. 
In the jittering jets explosion mechanism (which is within the general frame of the jet feedback mechanism; for a review see \citealt{Soker2016Rev}), on the other hand, stochastic angular momentum of the mass that the newly born NS or black hole accrete leads to the formation of an intermittent accretion disk, that it turn launches jittering jets. 
(e.g., \citealt{PapishSoker2011, GilkisSoker2014, GilkisSoker2015, Quataertetal2019}).
The jittering jets explosion mechanism does not require the pre-collapse core to rotate. 
  
In some cases we expect the morphology of the CCSN ejecta to have a large axisymmetrical departure from a spherical symmetry, and instead of two Ears to have two very large bipolar lobes. This might be the case when late energetic jets follow the explosion in the frame of the jet feedback explosion mechanism (e.g., \citealt{Gilkisetal2016Super}), when the last jet-launching episode of the jittering jets explosion mechanism launches energetic jets that inflate large bipolar lobes (or very large Ears), and in the neutrino driven mechanism if there is a massive enough fall back that is accreted through an accretion disk.  \cite{Stockingeretal2020} obtain an accretion disk in their simulations of neutrino driven explosions, but to form the large bipolar lobes we study here the fall back flow should be more massive. 
Another possibility is that a binary companion spins-up the pre-collapse envelope, leading to a bipolar explosion \citep{ChevalierSoker1989}. 

In a recent paper we described the effect that such a bipolar explosion might have on the light curve of a CCSN for an observer in the equatorial plane of the bipolar structure \citep{KaplanSoker2020b}. 
We built a geometrical toy model for the ejecta, that we describe here in sections \ref{sec:basics} and \ref{sec:model}. 
We found that an observer located in and near the equatorial plane will find a more rapid luminosity decline than in a spherical explosion, and even  an  abrupt luminosity drop.  
In this study we apply this toy model (section \ref{sec:model}) to the recently observed SN~2018gk (ASASSN-18am; \citealt{Boseetal2020}).
In section \ref{sec:fraction} we estimate the fraction of CCSNe that are bipolar and that we observe from their equatorial planes. 
We summarise our results in section \ref{sec:summary}.

\section{The basics of the bipolar toy model}
\label{sec:basics}

The details of the two-components bipolar toy model that leads to the geometrically modified light curve that we study here, and the justifications for using this toy model are in our earlier paper \citep{KaplanSoker2020b}. 
  In this section we present the basic assumptions and goals of the bipolar toy model, and in section \ref{subsec:general_model} we present the construction of the bipolar structure.

A full calculation of the light curve of a SN, in spherical and in non-spherical geometries, should include a rich variety of processes and values of physical variables (e.g., \citealt{EnsmanBurrows1992, Taddiaetal2016, Stritzingeretal2018}). Examples of these include the mass of newly synthesised radioactive isotopes and their mixing in the envelope (e.g., \citealt{Woosleyetal1994}). The mixing determines the escape of gamma rays from radioactive decay (e.g.,  \citealt{Berstenetal2012} ), which in turn implies that a full calculation should include the radiative transfer inside the ejecta in different electromagnetic bands (e.g., \citealt{Bufanoetal2014}).
Other effects are due to the shocks in the ejecta, including the shock breakout (e.g., \citealt{NakarPiro2014}). 
 
 The toy model circumvents the very complicated simulations that are required to calculate the light curve of non-spherical ejecta. We do so by assuming that in a spherical explosion the processes listed above lead to a light curve that has no rapid decline, but otherwise it is similar to the light curve we want to study. We take such a light curve to be the template light curve of the toy model. We then examine how the bipolar geometry might modify the light curve to have a rapid decline. In other words, we limit our study to the determination of the role of bipolar geometry. We also assume that the processes above combined to give a uniform temperature on the photosphere of the bipolar ejecta. The toy model includes a variation of the photospheric temperature with time, as found in spherical simulations (e.g. \citealt{Taddiaetal2016}).

 This prescription for reproducing the rapid decline in the light curve by the bipolar structure implies that for the template light curve we must take a light curve from a supernova that does not have a rapid decline, but that otherwise has a similar light curve. There is no free choice in the template light curve. In turn, this also implies that we assume that the explosion that leads to the template light curve is spherical (or has only a small departure from a spherical geometry). However, we do not know whether the light curve that we use here as a template (section \ref{subsec:Fitting}) comes from a spherical or non-spherical CCSN.

The omission of the calculations of the processes that we listed above comes with some drawbacks. For example, we do not take into account other effects that can lead to a rapid decline in the light curve. One such effect comes from a progenitor with an extended low-mass envelope around a compact core. This can lead to a rapid drop in the light curve (e.g., \citealt{NakarPiro2014}), as the expanding low mass gas allows gamma rays from radioactive decays to escape. For the same reason we are unable to rule out other models for a rapid decline in the light curve. We limit the goals of the present study to show that a bipolar ejecta might explain the rapid decline in the light curve of the type IIb CCSN SN~2018gk, and to determine some plausible parameters for such a bipolar explosion.

Another drawback is that we cannot determine the role that the bipolar geometry plays in the light curve for an observer far from the equatorial plane.

\section{A bipolar toy model for SN~2018gk}
\label{sec:model}
\subsection{The geometrically modified light curve}
\label{subsec:general_model}

We construct the ejecta from two components, the polar ejecta that is the ejecta within an angle of $\theta < \theta_0$ from the symmetry axis, and an equatorial ejecta that is in the section $\theta_0  \le \theta \le 180^\circ - \theta_0 $. Namely, within a section bounded by the angle of $90^\circ-\theta_0 $ from the equatorial plane and on both sides of the equatorial plane. 
We calculate the expansion of the polar ejecta (equatorial ejecta) as if it is a spherical shell of mass $M_{\rm po}$ ($M_{\rm eq}$) and an energy of $E_{\rm po}$ ($E_{\rm eq}$). The actual total mass and total energy of the polar ejecta in both sides of the equatorial plane, and those of the equatorial ejecta are 
\begin{eqnarray}
\begin{aligned}
& M_{\rm po,\theta_0}=(1-\cos \theta_0) M_{\rm po}; \quad  E_{\rm po,\theta_0}=(1-\cos \theta_0) E_{\rm po} 
\\ & 
M_{\rm eq,\theta_0}=\cos \theta_0 M_{\rm eq}; \quad  
E_{\rm eq,\theta_0}=\cos \theta_0 E_{\rm eq},
\label{eq:MEpotheta0}
\end{aligned}
\end{eqnarray}
 respectively.
 
The spherical expansion according to which each component expands is that of a spherical explosion according to \cite{ChevalierSoker1989} as in equations 1-6 of \cite{SuzukiMaeda2019}  with $\delta=1$ and $m=10$. The density profile is 
\begin{equation}
\rho = \begin{cases}
        K t^{-3} E^{-3/2} M^{5/2} \left( \frac{r}{t v_{\rm br}} \right)^{-1} 
        & r\leq t v_{\rm br}
        \\
        K t^{-3} E^{-3/2} M^{5/2} \left( \frac{r}{t v_{\rm br}} \right)^{-10} 
        & r>t v_{\rm br},
        \end{cases}
\label{eq:density_profile}
\end{equation}
where $K$ is a numerical constant and where the break in the density profile is at a velocity of $v_{\rm br}=1.69({E}/{M})^{1/2}$. We calculate the location of the photosphere to be where the optical depth to infinity is $\tau=2/3$ under the assumption of a constant opacity $\kappa$ (see equations 3-4 in \citealt{KaplanSoker2020b} for the location of the photosphere). 

We further assume that the entire photosphere, both of the polar component and of the equatorial component, has a uniform time-dependent effective temperature. The uniform temperature implies that the luminosity of ejecta at any given time is 
\begin{eqnarray}
\begin{aligned} 
L_{\rm eff} (t) = 4 \sigma A_{\rm cross}(t) T^4_{\rm eff}(t),
\end{aligned}
\label{eq:Leff}
\end{eqnarray}
where $A_{\rm cross}$ is the projection (cross section) of the photosphere on the plane of the sky, and $\sigma$ is the Stefan–Boltzmann constant. 

We do not calculate radiative transfer and for that cannot calculate the light curve directly. We rather use an observed light curve of another CCSN that has a similar light curve shape but that does not have a rapid drop after maximum  (section \ref{sec:basics}).   We take this template light curve to be, up to a scaled luminosity, the luminosity of a spherical CCSN, $L_{\rm sp,eq}(t)$, that has properties like the equatorial ejecta, namely total mass and energy of $M_{\rm eq}$ and $E_{\rm eq}$, respectively. 
 
The usage of the template light curve and equation (\ref{eq:Leff}) with a uniform effective temperature give the expression for the geometrically modified light curve of the non-spherical ejecta 
\begin{eqnarray}
L_{\rm SN}(t) = L_{\rm sp,eq} (t) \frac{A_{\rm cross}(t)}{A_{\rm sp,eq}(t)},  
\label{eq:lum}
\end{eqnarray}
where  $A_{\rm sp,eq}(t)$ is the projection (cross section) of the photosphere of a spherical ejecta on the plane of the sky. For the calculation of the ratio of the cross sections (ratio of projected areas) and then the light curves, we use MATLAB. 
  
Since we take the mass of the polar ejecta to be smaller than the equatorial mass, and its energy larger than that of the equatorial ejecta, the polar ejecta expands much faster. At early times the polar ejecta inflates a larger photosphere. This, by equation (\ref{eq:lum}), results in a brighter CCSN. The additional energy comes from the jets that inflate the CCSN. At later times the polar ejecta becomes optically thin and its photosphere becomes very small. This causes a rapid drop in luminosity for an observer in and near the equatorial plane, because now the equatorial ejecta hides the deep polar photosphere. Namely, for an equatorial observer at late times ${A_{\rm cross}(t)}< {A_{\rm sp,eq}(t)}$. We next demonstrate this geometry. 

\subsection{Evolution of the photosphere}
\label{subsec:photosphere}

\begin{figure}[ht!]
	\centering
	\includegraphics[trim=5.2cm 8.2cm 4cm 8.2cm ,clip, scale=0.8]{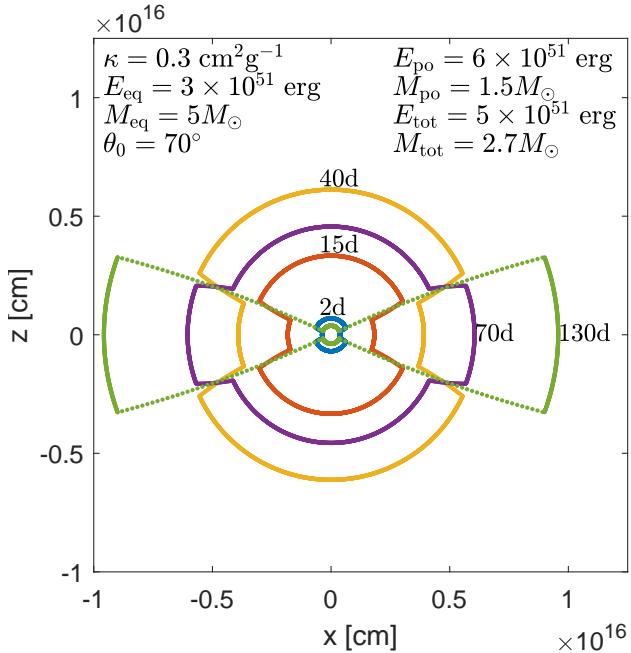}
	\caption{
The projection of the photosphere (cross section) of our two-components bipolar toy model for an equatorial observer at five times. Namely, the lines show the photosphere in the meridional plane at five times. We list the parameters of the model for this case in the figure. Each color represents the photosphere at a different time as we indicate in days. At early times (here of $2$, $15$ and $40 \days$) the photosphere of the polar ejecta expands faster than the photosphere of the equatorial ejecta. At later times (here of $70$ and $130 \days$) the outer parts of the polar ejecta becomes optically thin, and the photosphere in the polar directions rapidly recedes.
	}
	\label{fig:shapes}
\end{figure}

We present in Fig \ref{fig:shapes} the shape of the photosphere of the two-components bipolar toy model at five times. We calculate the photosphere, i.e., where the optical depth to infinity along a radial line is $\tau=2/3$, for the density profile of equation (\ref{eq:density_profile}), but separately for the polar and equatorial components (equations 3-4 in \citealt{KaplanSoker2020b}). The opacity in this case is $\kappa=0.3$ and the angle of separation between the two components is $\theta_0=70^{\circ}$. As we explain in section \ref{subsec:general_model}, each of the two ejecta components expand as if it is part of a spherical explosion. For the model we plot in Fig. \ref{fig:shapes} the masses and energies of the spherical explosions are $M_{\rm po}=1.5 M_{\odot}$, 
$E_{\rm po}=6 \times 10^{51}$, $M_{\rm eq}=5 M_{\odot}$, $E_{\rm eq}=3 \times 10^{51}$, respectively. 
The actual masses and energies are according to equation (\ref{eq:MEpotheta0}), such that the total energy and mass of our model with these parameters are $E_{\rm SN} = (1-\cos{\theta_0})E_{\rm po}+\cos{\theta_0}\;E_{\rm eq} = 5 \times 10^{51} \erg$, and $M_{\rm SN}= (1-\cos{\theta_0})M_{\rm po}+\cos{\theta_0}\;M_{\rm eq} = 2.7 M_\odot$. 
We note that \cite{Boseetal2020} estimate the mass and energy of the ejecta of SN~2018gk to be $M_{\rm 18gk} \simeq 3.4 M_\odot$ and $E_{\rm 18gk} \simeq (5-9) \times 10^{51} \erg$, respectively.  

Fig. \ref{fig:shapes} quantitatively presents the main property of the bipolar model. At early times the photosphere is larger in the polar directions (the gas within $\theta_0 =70^\circ$ from the symmetry axis $x=y=0$) because the polar ejecta moves faster. These are the blue line at $t=2 \days$, the red line at $t=15 \days$, and the yellow line at $t=40 \days$. The optical depth decreases faster in the polar regions, until it becomes optically thin in our toy model. Therefore, the photosphere moves inward in the polar ejecta, as we see by the purple line at $t=70 \days$. The cross section of the ejecta for an equatorial observer decreases, and so is the luminosity (eq. \ref{eq:lum}). At even later times the equatorial ejecta hides the polar photosphere from an equatorial observer, as the green line presents at $t=130 \days$. 

\subsection{Fitting the light curve of SN~2018gk}
\label{subsec:Fitting}

\begin{figure}[ht!]
	\centering
\includegraphics[trim=3.5cm 8cm 4cm 8cm ,clip, scale=0.6]{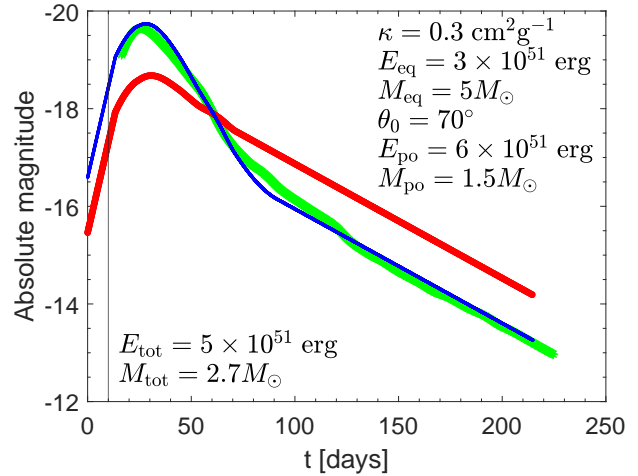}
\caption{Fitting the observed absolute magnitude in V band with the two-components bipolar toy model for an equatorial observer. 
The thick-green line is the observed  absolute magnitude in V band of SN~2018gk that we base on \cite{Boseetal2020}.
The thick-red line is the light curve of an explosion having the parameters we use for the equatorial ejecta, $L_{\rm sp,eq}$. We base the shape of the thick-red line on the observed absolute magnitude in V band of SASSN-15nx from \cite{Boseetal2018},  which we assume to be a spherical explosion. 
We linearly extended the observed light curve of SASSN-15nx to earlier times, left to the vertical line on the figure. We require it for the fitting by the toy model. 
The thin-blue line is the geometrically-modified light curves as we calculate from equation (\ref{eq:lum}).
}
\label{fig:modifiedLC}
\end{figure}

We Perform the steps that we described in section \ref{subsec:general_model} to find the best fit of our bipolar toy model to the light curve of SN~2018gk (SASSN-18am). 
We first smooth the light curve of SN~2018gk, the absolute magnitude in V band, as \cite{Boseetal2020} present in their figure 2. We present it by the thick-green line in Fig. \ref{fig:modifiedLC}. We assume that most of the emission near the maximum is in the V band. 
We next find a CCSN that has a similar shape to the light curve of SN~2018gk but without the rapid drop after maximum, as it is this rapid drop that we want to explain. From the comparison that \cite{Boseetal2020} make between SN~2018gk and others, we find that CCSN ASASSN-15nx has the appropriate light curve (\citealt{Boseetal2018}), $L_{\rm 15nx}(t)$. Namely, we take the template light curve of  what we assume to be  a spherical CCSN with the properties of the equatorial ejecta in our toy model to be $L_{\rm sp,eq}(t)=\beta L_{\rm 15nx}(t)$, where beta is a scaling factor.
We use the scaling factor as we do not expect the CCSN that serves as the template light curve to have the exact same luminosity as that of the equatorial ejecta in our toy model. 
      
 We emphasise the following points (see also section \ref{sec:basics}). (1) We assume that the explosion of ASASSN-15nx was spherical (or with only a small departure from a spherical structure). However, we do not know whether it was really spherical. (2)  Both SN~2018gk (ASASSN-18am) and ASASSN-15nx belong to a particular class of over-luminous CCSNe that have short rise time to peak luminosity (e.g., \citealt{Arcavietal2016, Boseetal2018}). That both SN~2018gk and ASASSN-15nx belong to the same class is in accord with the toy model that we apply here, as this toy model requires the template light curve to be similar in shape to that of the SN we study, beside not having the rapid decline. (3) The unique nature of the light curve of ASASSN-15nx, in having a near-perfect linear decline (\citealt{Boseetal2018}), is not a problem as the SN we study here is also a rare one. 
   
We draw $L_{\rm sp,eq}(t)$ by the thick-red line in Fig. \ref{fig:modifiedLC} for $\beta=0.48$. Namely, we moved the light curve of CCSN ASASSN-15nx, $L_{\rm 15nx}(t)$, vertically down by an absolute magnitude of $0.8$.
We note that in this figure the discovery time of ASASSN-15nx is at $t=10 \days$. We extended the light curve to early times, left to the vertical line on the figure, by a linear fit because to calculate the light curve we need to start shortly after the explosion. 
We emphasise that this continuation of the light curve does not influence our results and conclusions, since we fit only the peak, so the shape of the light curve before the peak is irrelevant to our model and serves only a technical purpose.
   
Once we have the template light curve we search the parameter space of our bipolar toy model (section \ref{subsec:general_model}) to find a good fit to the light curve of SN~2018gk. We find a set of parameters that yield a satisfactory fit to the light curve of SN~2018gk.
We presented the time evolution of the photosphere of this set of parameters in Fig. \ref{fig:shapes}, and in Fig. \ref{fig:modifiedLC} we present the light curve that we obtain with these parameters by the thin-blue line. 
The fit is not perfect, but we consider it adequate for the simple two-components bipolar toy model we apply. For example, the break in the light curve of SN~2018gk is about $42 \days$ from the peak, and in our model it is about $54 \days$ after the peak. 
       
 In finding the best fit that we present in Fig. \ref{fig:modifiedLC} we have tried tens of cases within the parameters range that the observation constrain. In particular, a Type IIb CCSN cannot have a too light (as there is some hydrogen) or a too heavy (as the hydrogen mass is very low) ejecta. Within this range the best fit we find is unique, although, for example, we could increase mass by up to $\approx 20\%$ and energy by up to about a factor of $\approx 2$ and still have a reasonable fit (not as good as the one in Fig 2, but not by much). We do not present the tens of cases we have tried, but rather limit this text to present the role of the different parameters of the toy model.

\begin{figure*} 
\centering
\begin{tabular}{cc}
\includegraphics[trim=3.6cm 8cm 4cm 9cm ,clip, scale=0.6]{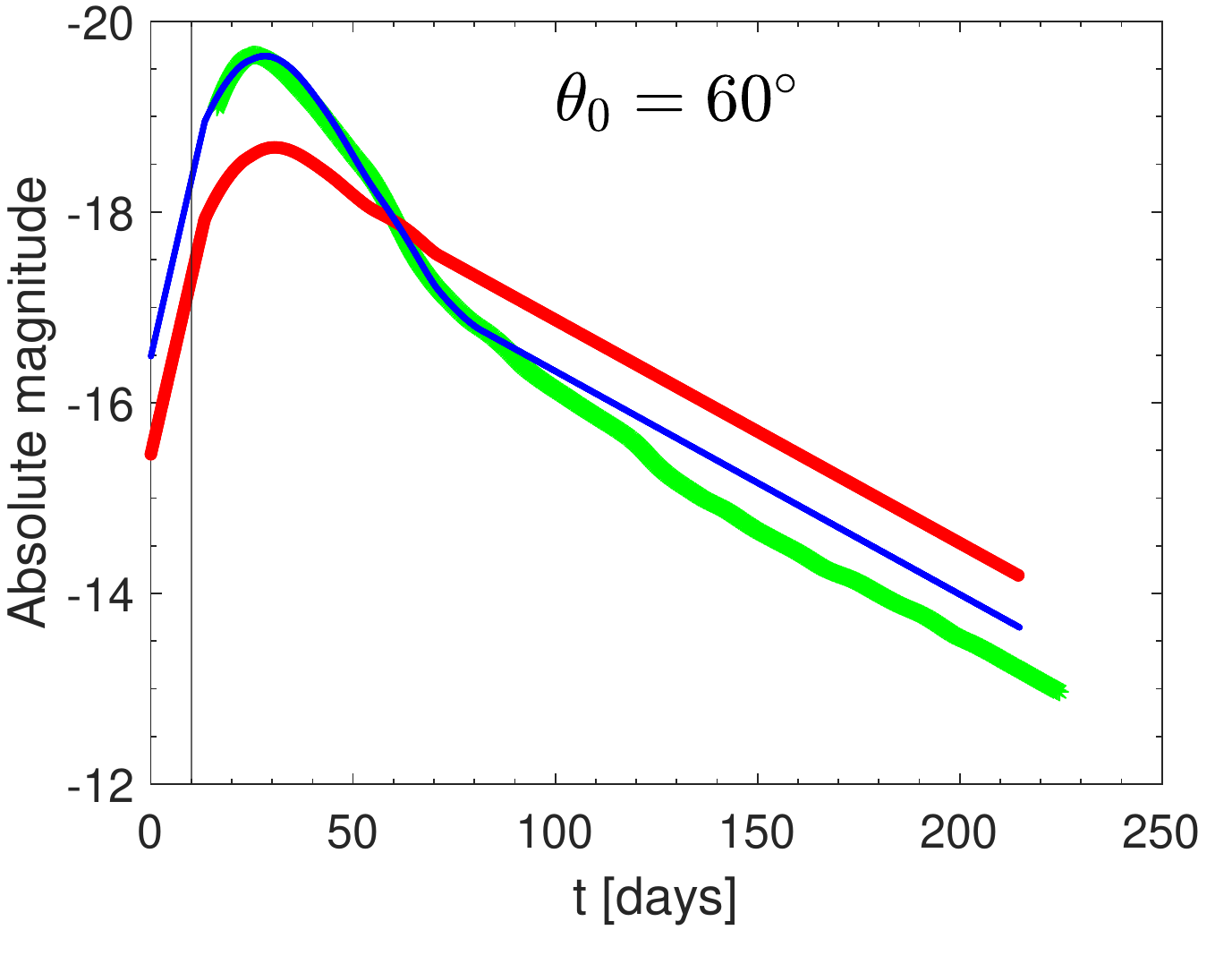}
\includegraphics[trim=3.6cm 8cm 4cm 9cm ,clip, scale=0.6]{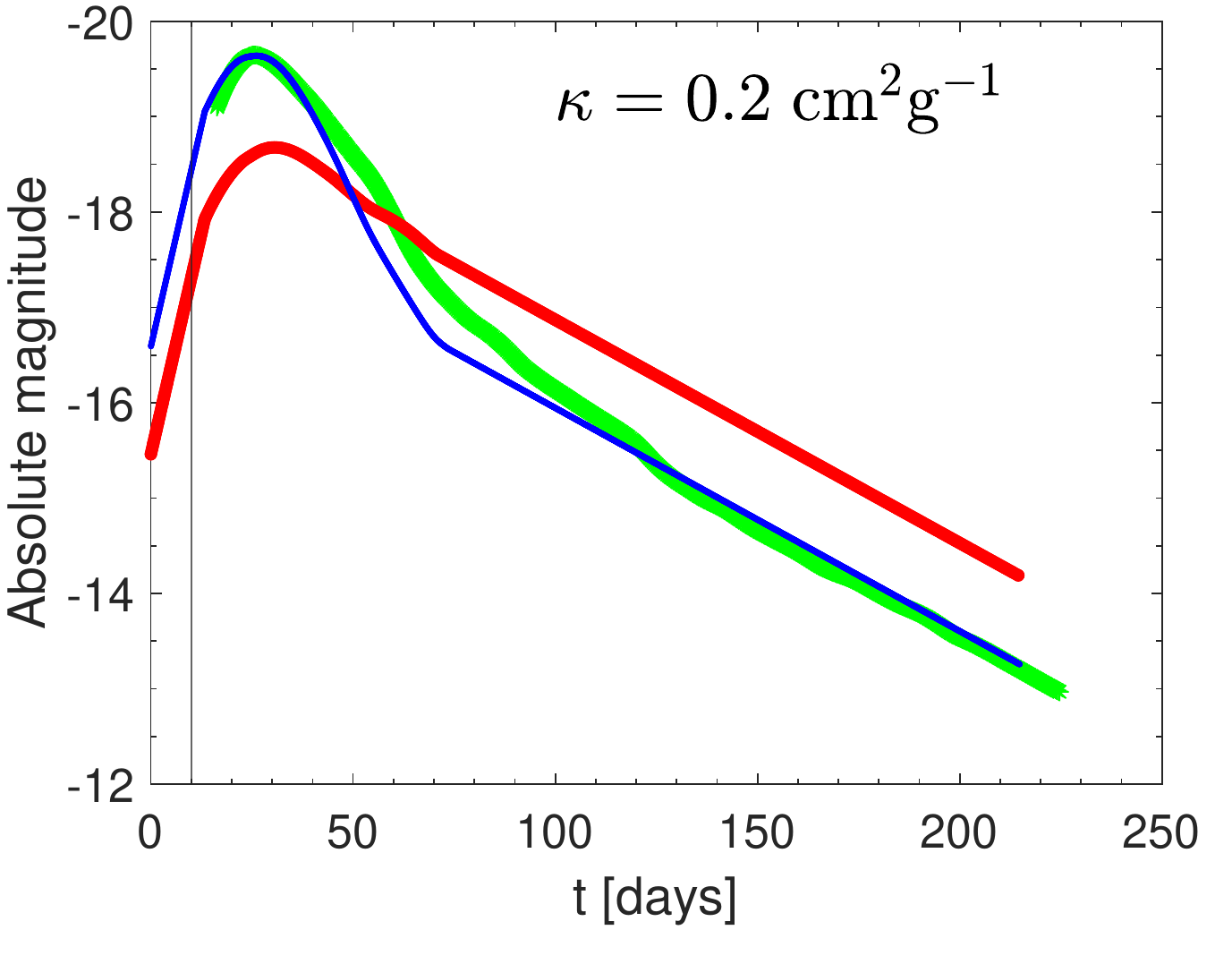} \\ 
\includegraphics[trim=3.6cm 8cm 4cm 9cm ,clip, scale=0.6]{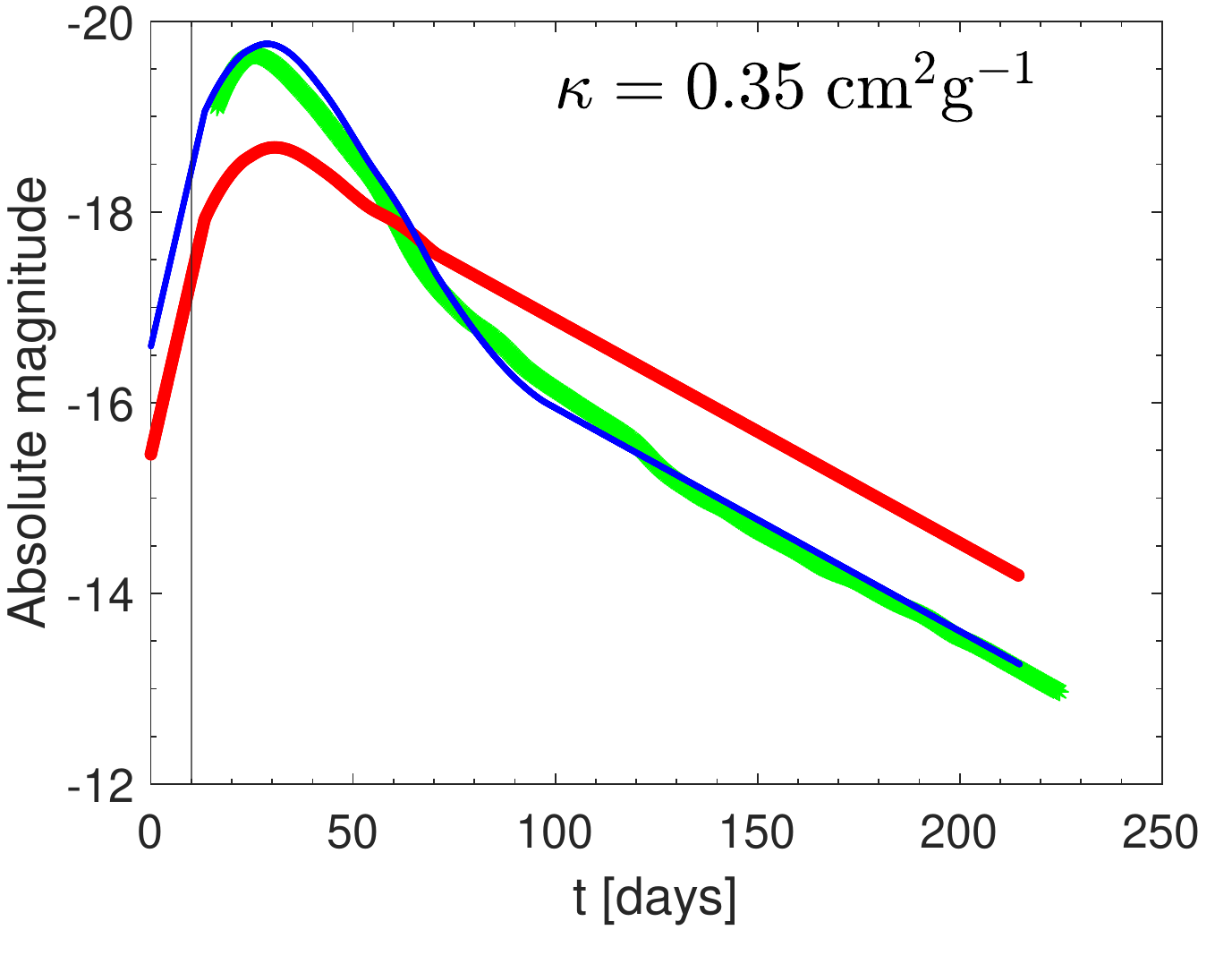}
\includegraphics[trim=3.6cm 8cm 4cm 9cm ,clip, scale=0.6]{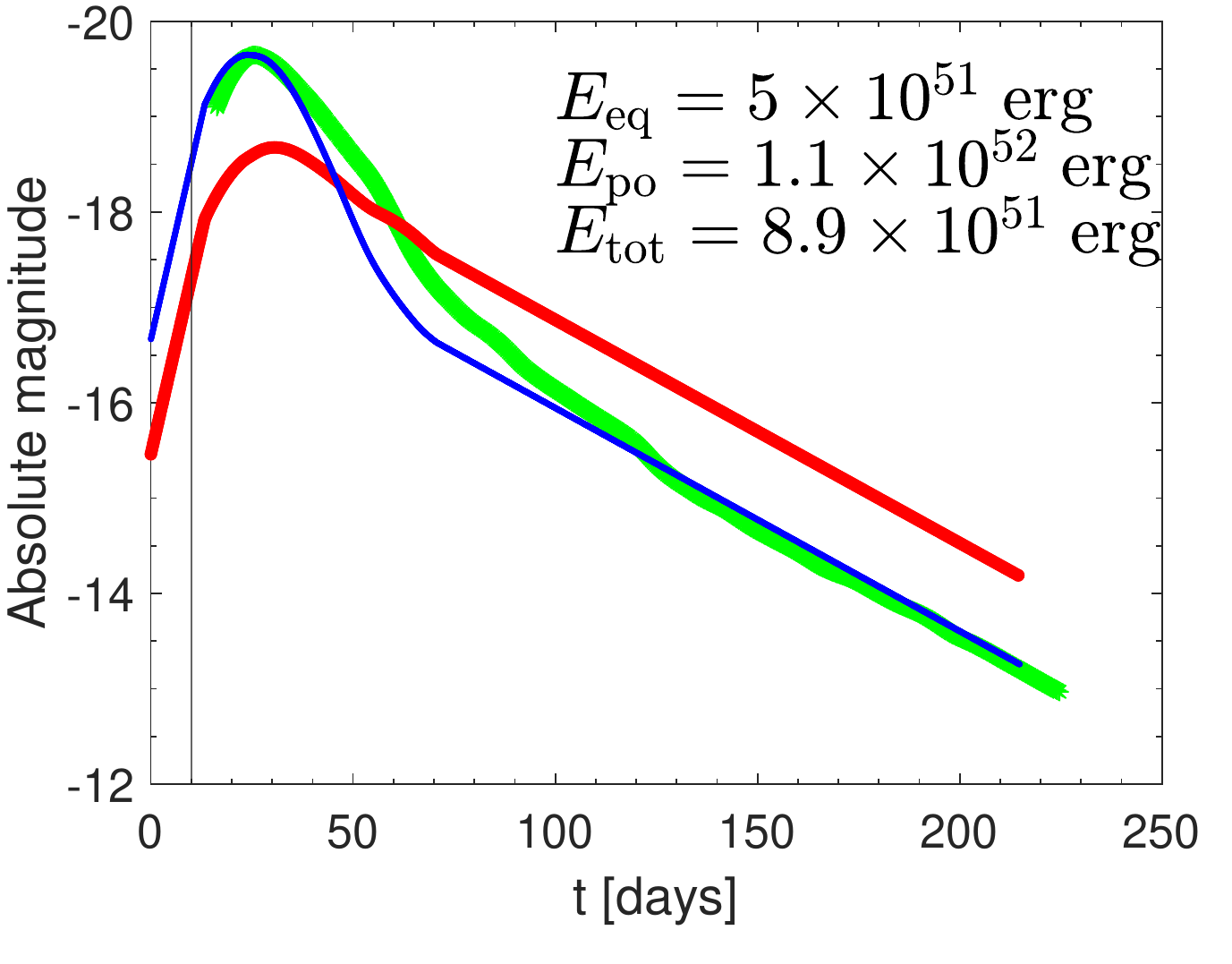} \\ 
\includegraphics[trim=3.6cm 8cm 4cm 9cm ,clip, scale=0.6]{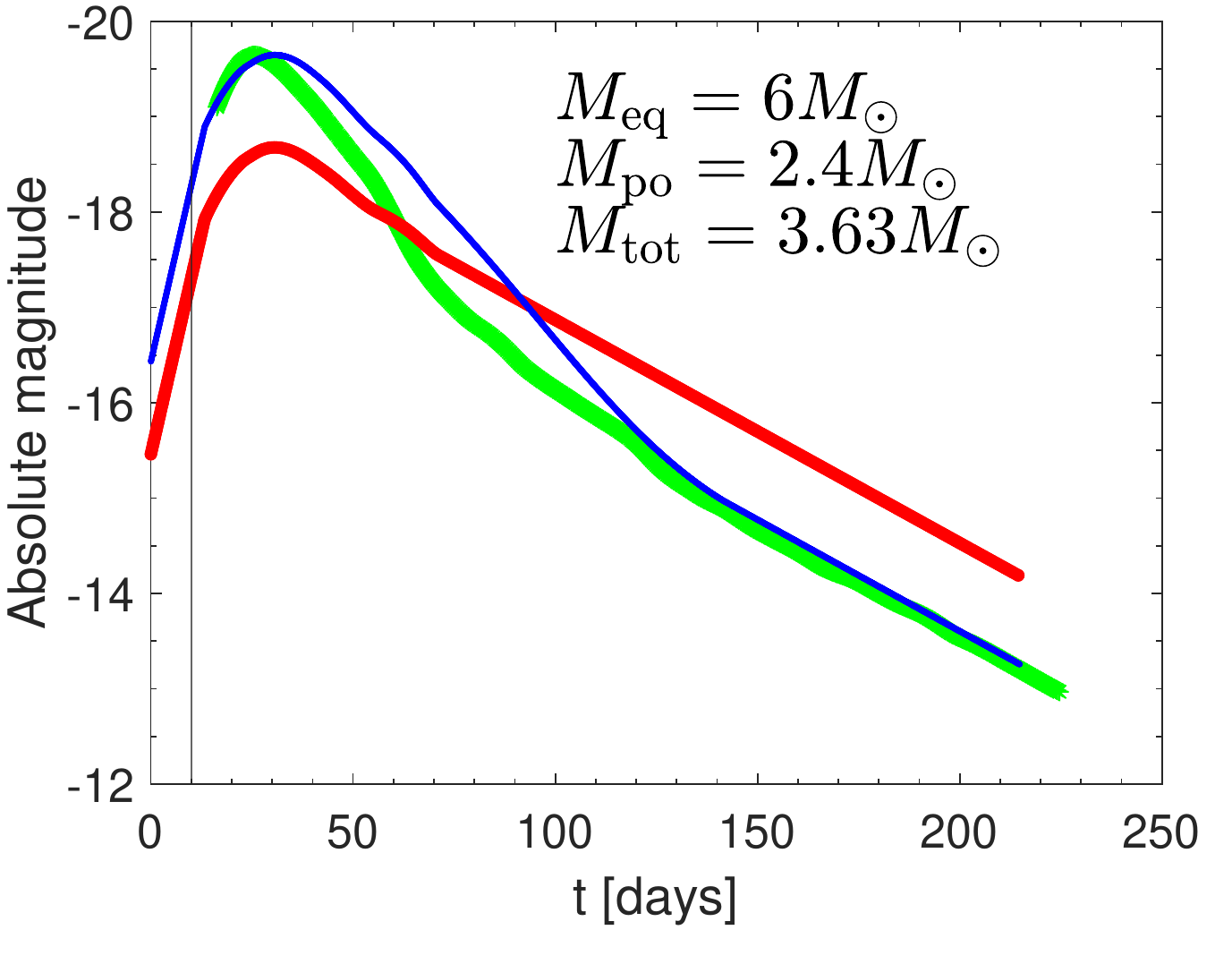}
\includegraphics[trim=3.6cm 8cm 4cm 9cm ,clip, scale=0.6]{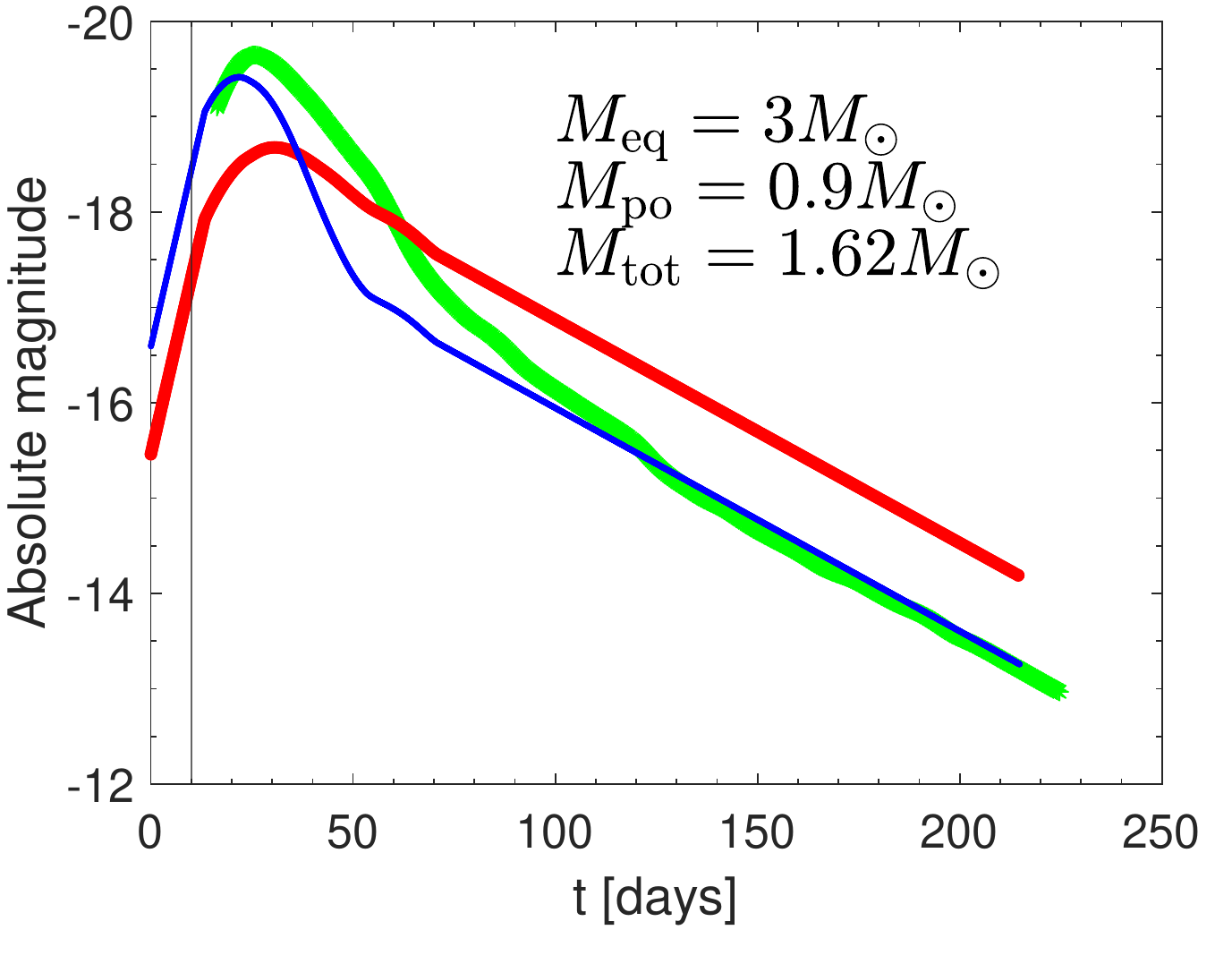} \\ 
\end{tabular}
\caption{The light curves as we describe in Fig. \ref{fig:modifiedLC}, but for different parameters of the two-components bipolar toy model (which affect only the thin-blue line). In each panel we list the parameter(s) that we vary with respect to those of Fig. \ref{fig:modifiedLC}. 
}
  \label{fig:Parameters}
    \end{figure*}

To explore the role of some parameters in the two-components bipolar toy model we calculate the light curves for six more cases and present their light curves by the thin-blue lines in Fig. \ref{fig:Parameters}. The inset of each panel lists the parameters that we vary with respect to the parameters of our best fit that we present in Fig. \ref{fig:modifiedLC}. All cases show a rapid drop, relative to the template light curve, and then a break back to a shalower decline. 

The conclusion of this section is that we can adequately fit the light curve of SN~2018gk with our simple two-components bipolar toy model. We further can constrain the parameters of this CCSN to have a total mas of $M_{\rm tot} \simeq 3 M_\odot$ and a total kinetic energy to $E_{\rm tot} \simeq 5 \times 10^{51} \erg$. These parameters are similar to those that \cite{Boseetal2020} estimate. 

The Neutrino driven explosion mechanism cannot reach such a high explosion energy, whether this mechanism can work or not for low energies (e.g., \citealt{Fryer2006, Fryeretal2012, Papishetal2015a, Sukhboldetal2016, SukhboldWoosley2016, Gogilashvilietal2020}). 
This leaves an energetic explosion that is driven by jets.
These jets are likely to shape the ejecta into a bipolar structure, as we demand for the bipolar model we use in the present study. Our usage of the bipolar toy model is compatible with the high explosion energy of SN~2018gk.

 We comment on our finding that the best bipolar model to fit the light curve of SN~2018gk has a wide polar ejecta, i.e., a half opening angle (measured from the symmetry axis) of $\theta_0 \simeq 70^\circ$. This does not necessarily imply that the jets that power the energetic polar ejecta are very wide. The polar ejecta are actually the two lobes that the two jets inflated. The jets, therefore, can be much narrower than $\theta_0 \simeq 70^\circ$. In a recent study \cite{AkashiSoker2020} conduct three-dimensional hydrodynamical simulations and show that jets with a half-opening angle of $20^\circ$ inflate two opposite and very wide bubbles, with half opening angles that reach $\approx 70^\circ$, despite that their numerical settings were for less energetic jets than what we need here.

\section{The fraction of equatorially observed bipolar CCSNe}
\label{sec:fraction}
 
We crudely estimate the fraction of observed CCSNe that might show a light curve with the property that we studied above. Namely, CCSNe that have a rapid drop from maximum light and then a break to a shallower decline, and which the model of a bipolar explosion with an equatorial observer might account for.

The last jets of an explosion driven by jittering jets in a progenitor with slowly rotating pre-collapse core might inflate `Ears'. However, in most cases (but maybe not in all) these `Ears' are too small (section \ref{sec:intro}) to account for a rapid drop in the light curve. 
According to the bipolar model the explosion should inflate large polar lobes. For that we require energetic jets or a flat rapidly rotating progenitor (section \ref{sec:intro}). Both of these cases to form a bipolar explosion require a binary companion to spin-up the progenitor. 

In the case of energetic jets that the newly born NS (or black hole) launches the companion should spin-up the core, such that the accreted gas has high specific angular momentum that maintains the jets along a fixed-axis. This requires that the companion spirals-in inside the progenitor envelope all the way to influence the core, e.g., by tidal interaction. The companion likely removes most or all of the hydrogen-rich envelope to form a stripped-envelope CCSN. Namely, the CCSN is a SN IIb, a SN Ib or a SN Ic (SNe Ibc). SN~2018don that we modelled in a previous paper \citep{KaplanSoker2020b} and that has a light curve with an abrupt drop \citep{Lunnan2020}, is a  type Ic SN. The SN that we study here, SN~2018gk is a Type IIb CCSN. 

The fraction of stripped-envelope CCSNe (SNe Ib + SNe Ic + SNe IIb) increase on average with the host galaxy mass (e.g., \citealt{Grauretal2017a, Grauretal2017b}). Approximately, the fraction of stripped-envelope CCSNe out of all CCSNe is $f_{\rm SE} \simeq 0.3$ \citep{Grauretal2017b}. 

Only a fraction $f_{\rm s, SE}$ of stripped-envelope CCSNe has a binary interaction that spins-up the core such that it launches energetic jets to form a bipolar explosion. This fraction is the most uncertain factor in our estimate. We expect that this fraction is relatively large for SNe Ic that seem to require a binary companion, $f_{\rm s,Ic} \approx 0.5$, and smaller for SNe Ib and SNe IIb that maintain a massive helium-rich layer, $f_{\rm s,b} \approx 0.1-0.3$. But these are more of a guess than a calculation.  

From our two-components bipolar toy model in \cite{KaplanSoker2020b} and in section \ref{sec:model} here, the observer should be close to the orbital plane, within $\alpha_O \la 20^\circ$ so that the equatorial ejecta hides the polar photosphere at late times. This means that only a fraction of $f_{\rm O} =  \sin \alpha_O \simeq 0.35$ of observers can observe the effect we study here. 
Overall, the fraction of stripped-envelope CCSNe that we might observe to have a rapid decline and then a break because of a bipolar structure out of all CCSNe is
\begin{equation}
f_{\rm SE,CC} \simeq f_{\rm SE} f_{\rm SE,SE} = 
f_{\rm SE} f_{\rm s,SE} f_{\rm O} \approx  0.015-0.05
\label{eq:fractions}
\end{equation}
where $f_{\rm SE, SE} \approx 0.05-0.15$ is the fraction of these systems out of stripped-envelope CCSNe. In some of these rapid decline CCSNe the effect will be too small to notice, or might be smeared by other effects, such as interaction with a pre-explosion circumstellar matter (CSM).  

Let us consider the case of SN II and SNe IIb where there is an extended hydrogen-rich envelope. In these cases a flatten envelope due to a common envelope evolution (section \ref{sec:intro}) might also lead to the formation of a dense and a slow equatorial ejecta and a faster polar ejecta. In these cases, though, interaction with a CSM might obscure the rapid declines. We expect such systems to contribute a small fraction to the systems we study here. 

Overall, our crude estimate is that the fraction of all CCSNe that show a rapid decline in light curve due to a bipolar explosion that we observe from the equatorial plane is 
$f \approx 2 - 5\%$ of all CCSNe.

\section{Summary}
\label{sec:summary}
     
We extended our study on the influence of a bipolar CCSN ejecta on the light curve that an equatorial observer measures, what we termed a ``geometrically modified light curves'' \citep{KaplanSoker2020b}. In most cases we attribute the bipolar morphology to energetic jets that drive the explosion and shape the ejecta. We built a two-components bipolar toy model (Fig. \ref{fig:shapes}) to estimate the modifications on the light curve. We do not calculate the light curve, but rather use a template light curve that we base on a different CCSN that is similar to the target CCSN, but that has no rapid drop in its light curve. 
Overall, the bipolar morphology increases the maximum luminosity and causes a steeper decline for an equatorial observer, relative to a similar spherical explosion (for more details of the toy model see \citealt{KaplanSoker2020b}). 

In the first study \citep{KaplanSoker2020b} we applied the two-components bipolar toy model to SN~2018don, a SN Ic that has an abrupt drop in its light curve \citep{Lunnan2020}. In this study we applied the toy model to a SN IIb, namely, SN~2018gk with a rapid decline from maximum luminosity (\citealt{Boseetal2020}). We could find a good fit to this rapid decline with a break to a shallower decline at a late time with our toy model (Fig. \ref{fig:modifiedLC}). In Fig. \ref{fig:Parameters} we explored the influence of some parameters in our toy model. 

The total energy and mass of our toy model for SN~2018gk are $E_{\rm SN} = 5 \times 10^{51} \erg$ and $M_{\rm SN}= 2.7 M_\odot$, respectively. These are similar to the values that  \cite{Boseetal2020} estimate. The main issue here is that the explosion energy of SN~2018gk is $E > 3 \times 10^{51} \erg$ is more than what a neutrino-driven explosion mechanism can explain. We take it to imply that energetic jets exploded this CCSN. Such jets must shape the ejecta in a way that might influence the light curve, that we examined for equatorial observers. 

We crudely estimate that $f \approx 2-5 \%$ of all CCSNe show this behavior (section \ref{sec:fraction}). Most of these are stripped-envelope CCSNe. 
  To observe this rapid decline the observer should be within an angle of 
$\alpha_O \la 20^\circ$ from the equatorial plane. The probability to be within this angle is about $34 \%$. Namely, we crudely estimate that about $f_{\rm bip} \approx 5-15 \%$ of all CCSNe have a pronounced bipolar morphology. 

The main point that our present study emphasises is that any analysis of luminous CCSNe should pay attention to bipolar ejecta morphologies.
On a broader scope, this study adds to the rich variety of effects that jets might have during and after the explosion of CCSNe.

Finally, we note that \cite{Gutierrezetal2020} argue that the fast-declining CCSN  SN~2017ivv had an aspherical explosion. Our model can in principle account for the fast-declining light curve for this aspherical CCSN if we observe it from near the equatorial plane. 
 
The next step in this line of study is to conduct two-dimensional or three-dimensional hydrodynamical simulations of the jet-shaped ejecta (e.g., \citealt{AkashiSoker2020} for late jets rather than at explosion), and to include radiative transfer. Such simulations will show the light curve from all directions. 

\section*{Acknowledgments}
We thank Ari Laor  and an anonymous referee  for useful comments.  This research was supported by a grant from the Israel Science Foundation (420/16 and 769/20) and a grant from the Asher Space Research Fund at the Technion.


\label{lastpage}

\begin{thebibliography}{}

\bibitem[Akashi et al.(2018)]{Akashietal2018} Akashi, M., Bear, E., \& Soker, N.\ 2018, \mnras, 475, 4794

\bibitem[Akashi \& Soker(2020)]{AkashiSoker2020} Akashi, M. \& Soker, N.\ 2020, arXiv:2006.01717


\bibitem[Aloy et al.(2000)]{Aloyetal2000} Aloy M. A., Muller E., Ibanez J.~M., Marti, J.~M., \& MacFadyen A.\ 2000, \apj, 531, L119
     
\bibitem[Arcavi et al.(2016)]{Arcavietal2016} Arcavi, I., Wolf, W.~M., Howell, D.~A., et al.\ 2016, \apj, 819, 35. doi:10.3847/0004-637X/819/1/35
   
\bibitem[Bear et al.(2017)]{Bearetal2017} Bear, E., Grichener, A., \& Soker, N.\ 2017, \mnras, 472, 1770
	
\bibitem[Bear, \& Soker(2018)]{BearSoker2018} Bear, E., \& Soker, N.\ 2018, \mnras, 478, 682

\bibitem[Bersten et al.(2012)]{Berstenetal2012} Bersten, M.~C., Benvenuto, O.~G., Nomoto, K., et al.\ 2012, \apj, 757, 31. doi:10.1088/0004-637X/757/1/31

\bibitem[Bose et al.(2018)]{Boseetal2018} Bose, S., Dong, S., Kochanek, C.~S., et al.\ 2018, \apj, 862, 107

\bibitem[Bose et al.(2019)]{Boseetal2019} Bose S., et al., 2019, ApJ, 873, L3

\bibitem[Bose et al.(2020)]{Boseetal2020} Bose, S., Dong, S., Kochanek, C.~S., et al.\ 2020, arXiv e-prints, arXiv:2007.00008

\bibitem[Bromberg \& Tchekhovskoy(2016)]{BrombergTchekhovskoy2016} Bromberg, O., \& Tchekhovskoy, A.\ 2016, \mnras, 456, 1739
	
\bibitem[Bromberg et al.(2014)]{Bromberg_jet} Bromberg, O., Granot, J., Lyubarsky, Y., et al.\ 2014, \mnras, 443, 1532.

\bibitem[Bufano et al.(2014)]{Bufanoetal2014} Bufano, F., Pignata, G., Bersten, M., et al.\ 2014, \mnras, 439, 1807. doi:10.1093/mnras/stu065

\bibitem[Burrows et al.(2007)]{Burrows2007} Burrows, A., Dessart, L., Livne, E., Ott, C.~D., \& Murphy, J.\ 2007, \apj, 664, 416

\bibitem[Chevalier, \& Soker(1989)]{ChevalierSoker1989} Chevalier, R.~A., \& Soker, N.\ 1989, \apj, 341, 867

\bibitem[Ensman \& Burrows(1992)]{EnsmanBurrows1992} Ensman, L. \& Burrows, A.\ 1992, \apj, 393, 742. doi:10.1086/171542

\bibitem[Feng et al.(2018)]{Fengetal2018} Feng, E.-H., Shen, R.-F., \& Lin, W.-P.\ 2018, \apj, 867, 130

\bibitem[Fryer(2006)]{Fryer2006} Fryer, C.~L.\ 2006, \nar, 50, 492

\bibitem[Fryer et al.(2012)]{Fryeretal2012} Fryer, C.~L., Belczynski, K., Wiktorowicz, G., Dominik, M., Kalogera, V., \& Holz, D.~E.\ 2012, \apj, 749, 91

\bibitem[Garc{\'{\i}}a et al.(2017)]{Garciaetal2017} Garc{\'{\i}}a, F., Su{\'a}rez, A.~E., Miceli, M., Bocchino, F., Combi, J.~A., Orlando, S., \& Sasaki, M.\ 2017, \aap, 604, L5

\bibitem[Gilkis(2018)]{Gilkis2018} Gilkis, A.\ 2018, \mnras, 474, 2419

\bibitem[Gilkis \& Soker(2014)]{GilkisSoker2014}  Gilkis, A., \& Soker, N.\ 2014, \mnras, 439, 4011

\bibitem[Gilkis \& Soker(2015)]{GilkisSoker2015}  Gilkis, A., \& Soker, N.\ 2015, \apj, 806, 2

\bibitem[Gilkis et al.(2016)]{Gilkisetal2016Super} Gilkis, A., Soker, N., \& Papish, O.\ 2016, \apj, 826, 178

\bibitem[Gogilashvili et al.(2020)]{Gogilashvilietal2020} Gogilashvili, M., Murphy, J.~W., \& Mabanta, Q.\ 2020, arXiv e-prints, arXiv:2007.06087

\bibitem[Gonz{\'a}lez-Casanova et al.(2014)]{Gonzalezetal2014} Gonz{\'a}lez-Casanova, D.~F., De Colle, F., Ramirez-Ruiz, E., \& Lopez, L.~A.\ 2014, \apjl, 781, L26

\bibitem[Graur et al.(2017a)]{Grauretal2017a} Graur, O., Bianco, F.~B., Huang, S., et al.\ 2017a, \apj, 837, 120

\bibitem[Graur et al.(2017b)]{Grauretal2017b} Graur, O., Bianco, F.~B., Modjaz, M., Shivvers, I., Filippenko, A.~V., Li, W., \& Smith, N.,\ 2017b, \apj, 837, 121

\bibitem[Grichener, \& Soker(2017)]{GrichenerSoker2017} Grichener, A., \& Soker, N.\ 2017, \mnras, 468, 1226

\bibitem[Guti{\'e}rrez et al.(2020)]{Gutierrezetal2020}  Guti{\'e}rrez, C.~P., Pastorello, A., Jerkstrand, A., et al.\ 2020, arXiv:2008.09628 

\bibitem[H{\"o}flich et al.(2001)]{Hoflich2001} H{\"o}flich, P., Khokhlov, A., \& Wang, L.\ 2001, 20th Texas Symposium on relativistic astrophysics, 586, 459
	
\bibitem[Inserra et al.(2016)]{Inserraetal2016}  Inserra, C., Bulla, M., Sim, S.~A., \& Smartt, S.~J.\ 2016, \apj, 831, 79

\bibitem[Kaplan \& Soker(2020a)]{KaplanSoker2020} Kaplan, N., \& Soker, N.\ 2020a, \mnras, 492, 3013

\bibitem[Kaplan \& Soker(2020b)]{KaplanSoker2020b} Kaplan, N., \& Soker, N.\ 2020b, \mnras, 494, 5909


\bibitem[Khokhlov et al.(1999)]{Khokhlovetal1999} Khokhlov, A.~M., H{\"o}flich, P.~A., Oran, E.~S., et al.\ 1999, \apjl, 524, L107
	

\bibitem[Lazzati et al.(2012)]{Lazzati2012} Lazzati, D., Morsony, B.~J., Blackwell, C.~H., \& Begelman, M.~C.\ 2012, \apj, 750, 68

\bibitem[Lopez \& Fesen(2018)]{LopezFesen2018} Lopez, L.~A., \& Fesen, R.~A.\ 2018, \ssr, 214, \#44
	
\bibitem[L{\'o}pez-C{\'a}mara et al.(2016)]{LopezCamaraetal2016}  L{\'o}pez-C{\'a}mara, D., Lazzati, D., \& Morsony, B.~J.\ 2016, \apj, 826, 180
	
\bibitem[L{\'o}pez-C{\'a}mara et al.(2014)]{LopezCamaraetal2014}  L{\'o}pez-C{\'a}mara, D., Morsony, B.~J., \& Lazzati, D.\ 2014, \mnras, 442, 2202

\bibitem[Lunnan et al.(2020)]{Lunnan2020} Lunnan, R., Yan, L., Perley, D.~A., et al.\ 2020, arXiv e-prints, arXiv:1910.02968

\bibitem[Maeda et al.(2012)]{Maedaetal2012} Maeda, K., Moriya, T., Kawabata, K., et al.\ 2012, \memsai, 83, 264
	
\bibitem[Margutti et al.(2014)]{Marguttietal2014} Margutti, R., Milisavljevic, D., Soderberg, A.~M., et al.\ 2014, \apj, 797, 107
	
\bibitem[Mauerhan et al.(2017)]{Mauerhanetal2017} Mauerhan, J.~C., Van Dyk, S.~D., Johansson, J., Hu, M., Fox, O.~D., Wang, L., Graham, M.~L., Filippenko, A.~V., \& Shivvers, I.\ 2017, \apj, 834, 118
	
\bibitem[Maund et al.(2007)]{Maundetal2007} Maund, J.~R., Wheeler, J.~C., Patat, F.,  Baade, D., Wang, L., H\"oflich, P.\ 2007, \mnras, 381, 201

\bibitem[Milisavljevic et al.(2013)]{Milisavljevic2013} Milisavljevic, D., Soderberg, A.~M., Margutti, R., et al.\ 2013, \apjl, 770, LL38
	
\bibitem[M{\"o}sta et al.(2014)]{Mostaetal2014} M{\"o}sta, P., Richers, S., Ott, C.~D., et al.\ 2014, \apjl, 785, L29

\bibitem[Nagakura et al(2011)]{Nagakuraetal2011} Nagakura H., Ito H., Kiuchi K., \& Yamada S.,\ 2011, ApJ, 731, 80

\bibitem[Nakar \& Piro(2014)]{NakarPiro2014} Nakar, E. \& Piro, A.~L.\ 2014, \apj, 788, 193. doi:10.1088/0004-637X/788/2/193

\bibitem[Nishimura et al.(2017)]{Nishimuraetal2017} Nishimura, N., Sawai, H., Takiwaki, T., Yamada, S., \& Thielemann, F.-K.\ 2017, \apjl, 836, L21
	
\bibitem[Orlando et al.(2016)]{Orlandoetal2016} Orlando S., Miceli M., Pumo M.~L., Bocchino F., 2016, ApJ, 822, 22

\bibitem[Papish et al.(2015)]{Papishetal2015a} Papish, O., Nordhaus, J., \& Soker, N.\ 2015, \mnras, 448, 2362

\bibitem[Papish \& Soker(2011)]{PapishSoker2011} Papish, O., \& Soker, N.\ 2011, \mnras, 416, 1697


\bibitem[Quataert et al.(2019)]{Quataertetal2019} Quataert, E., Lecoanet, D., \& Coughlin, E.~R.\ 2019, \mnras, 485, L83

\bibitem[Soker(2016)]{Soker2016Rev} Soker, N.\ 2016, \nar, 75, 1

\bibitem[Stockinger et al.(2020)]{Stockingeretal2020} Stockinger, G., Janka, H.-T., Kresse, D., et al.\ 2020,  \mnras, 496, 2039

\bibitem[Stritzinger et al.(2018)]{Stritzingeretal2018} Stritzinger, M.~D., Taddia, F., Burns, C.~R., et al.\ 2018, \aap, 609, A135. doi:10.1051/0004-6361/201730843

\bibitem[Sukhbold et al.(2016)]{Sukhboldetal2016} Sukhbold, T., Ertl, T., Woosley, S.~E., Brown, J.~M., \& Janka, H.-T.\  2016, \apj, 821, 38

\bibitem[Sukhbold \& Woosley(2016)]{SukhboldWoosley2016} Sukhbold, T., \& Woosley, S.~E.\ 2016, \apjl, 820, L38


\bibitem[Suzuki \& Maeda(2019)]{SuzukiMaeda2019} Suzuki, A., \& Maeda, K.\ 2019, \apj, 880, 150

\bibitem[Taddia et al.(2016)]{Taddiaetal2016} Taddia, F., Sollerman, J., Fremling, C., et al.\ 2016, \aap, 588, A5. doi:10.1051/0004-6361/201527811

\bibitem[Takiwaki \& Kotake(2011)]{TakiwakiKotake2011} Takiwaki, T., \& Kotake, K.\ 2011, \apj, 743, 30
	
\bibitem[Wang et al.(2001)]{Wangetal2001} Wang, L., Howell, D.~A., H{\"o}flich, P., \& Wheeler, J.~C.\ 2001, \apj, 550, 1030
		
\bibitem[Woosley et al.(1994)]{Woosleyetal1994} Woosley, S.~E., Eastman, R.~G., Weaver, T.~A., \& Pinto, P.~A., \ 1994, \apj, 429, 300. doi:10.1086/174319
     
\end{thebibliography}
\end{document}